\documentclass[prl,twocolumn,aps,showpacs,nofootinbib,floatfix,nobibnotes]{revtex4-2}
\usepackage{amsmath, bm}
\usepackage{amssymb}
\usepackage{graphicx}
\usepackage{verbatim}
\usepackage[colorlinks=true,linkcolor=blue,citecolor=blue,urlcolor=blue]{hyperref}
\usepackage{txfonts}
\usepackage{dsfont}
\usepackage{enumitem}
\usepackage[utf8]{inputenc}
\usepackage{bbold}
\usepackage{mathtools}
\usepackage{lipsum, babel}

\begin{document}

\title{Nonlinear Interferometry for Quantum-Enhanced Measurements of Multiphoton Absorption}
\author{Shahram Panahiyan$^{1,2,3}$, Carlos S\'anchez~Mu{\~n}oz$^4$, Maria V. Chekhova$^{5,6}$, and Frank Schlawin$^{1,2,3}$}
\email{shahram.panahiyan@mpsd.mpg.de}
\email{frank.schlawin@mpsd.mpg.de}
\affiliation{$^1$ Max Planck Institute for the Structure and Dynamics of Matter, Luruper Chaussee 149, 22761 Hamburg, Germany }
\affiliation{$^2$ The Hamburg Centre for Ultrafast Imaging, Luruper Chaussee 149, Hamburg D-22761, Germany}
\affiliation{$^3$ University of Hamburg, Luruper Chaussee 149, Hamburg, Germany}
\affiliation{$^4$ Departamento de F\'isica Te\'orica de la Materia Condensada and Condensed Matter Physics Center (IFIMAC), Universidad Aut\'onoma de Madrid, Madrid, Spain}
\affiliation{$^5$ Max-Planck Institute for the Science of Light, Staudtstr. 2, Erlangen D-91058, Germany}
\affiliation{$^6$ University of Erlangen-Nuremberg, Staudtstr. 7/B2, Erlangen D-91058, Germany}

\date{\today}

\begin{abstract}

Multiphoton absorption is of vital importance in many spectroscopic, microscopic or lithographic
applications. However, given that it is an inherently weak process, the detection of multiphoton absorption signals typically requires large field intensities, hindering its applicability in many practical situations.
In this work, we show that placing a multiphoton absorbent inside an imbalanced nonlinear interferometer can
enhance the precision of multiphoton cross-section estimation with respect to strategies based on direct transmission measurements by coherent or even squeezed light.
In particular, the power scaling of the sensitivity with photon flux can be increased by an order of magnitude compared to transmission measurements of the sample with coherent light, meaning that a signal could be observed at substantially reduced excitation intensities. 
Furthermore, we show that this enhanced measurement precision is robust against experimental imperfections leading to photon losses, which usually tend to degrade the detection sensitivity. We trace the origin of this enhancement to an optimal degree of squeezing which has to be generated in a nonlinear SU(1,1)-interferometer.

\end{abstract}

\maketitle

\emph{Introduction.---} Multiphoton absorption (mPA) is a nonlinear process in which several photons are simultaneously absorbed by the sample \cite{CRONSTRAND2005,Fabio2006,Rumi2010}. 
Large penetration depths and the nonlinear dependence on the beam profile, causing most of the signal to be generated in the confined area of maximal beam intensity, make it an appealing process for a variety of technological applications. 
Most famously, in nonlinear imaging, multiphoton processes can surpass the single-photon diffraction limit and thereby enhance the spatial resolution \cite{Klar2000, Brettschneider2007, So00}, pushing the resolution of optical microscopy to molecular scales.
mPA also forms the foundation for diverse applications ranging from
3D microfabrication \cite{Sun2000}, to optical data storage \cite{A.1989}, spectroscopy and microscopy~\cite{Helmchen2005,D2TC00191H,lin2012multiphoton} 
and even medical applications such as photodynamic therapy \cite{Atif2007}.
These advantages offered by mPA are, however, limited by the inherent weakness of nonlinear light-matter interactions \cite{Mollow,Boyd} and the resulting, very small multiphoton absorption cross sections, such that typically the use of strong, ultrafast lasers is the only way to overcome this problem and generate a measurable signal.

One way to circumvent this restriction may lie in the exploitation of quantum properties of light. 
In particular, two-photon absorption of entangled photons has gained prominence in this regard~\cite{Matthews2016,Dorfman16,JPhysB17,Villabona17, AccChemRes,Gilaberte2019,Szoke2020,Eshun2022,Yuanyuan2021}. It is known to scale linearly, rather than quadratically, with the photon flux~\cite{Klyshko1982, Georgiades95, Georgiades97, Javainen90}, and should lead to enhanced nonlinear responses. Early reports of very large enhancements (by many orders of magnitude) generated enormous interest in the field~\cite{Eshun2022}.
The strength of this enhancement is, however, the subject of intense current debate~\cite{Raymer2021, Burdick2021, Landes2021, Landes2021b, Parzuchowski2021,Hickam2022}. Moreover, even if large enhancements due to entanglement were feasible, due to its occurence in the few-photon regime, where the mean photon number per mode must be small, $\langle \hat{n} \rangle \lesssim 1$, it is not suitable for many practical applications, such as nonlinear imaging, where a nonlinear photon flux dependence, which requires $\langle \hat{n} \rangle \gg 1$, is crucial to enhance the resolution. Strong frequency correlations, which lie at the heart of the suspected advantage of entangled photon absorption, can still be observed even up to macroscopic regimes with $\langle \hat{n} \rangle \sim 10^8$~\cite{Cutipa2022}.

This motivates us to investigate, more broadly, the mPA of nonclassical states of light with mesoscopic character, i.e. with large photon numbers compared to entangled photon sources, but much weaker than high-intensity laser pulses.
With notable exceptions~\cite{Birchall20, Atkinson2021}, this photon number regime has received much less attention in the literature to date, even though it is very attractive for said applications. 
The rate of m-photon absorption scales with the m-th order correlation function, which, for bunched sources, grows exponentially with 
$m$~\cite{Spasibko2017}. But for practical applications, this enhancement has to be compared with the increased noise levels of these sources that may erode the benefits. 
We investigated two-photon absorption recently~\cite{Carlos2021, Panahiyan2022}, and found an improved sensitivity scaling for measurements of the squeezed quadrature. 
\begin{figure*}
\centering
\includegraphics[width=0.45\textwidth]{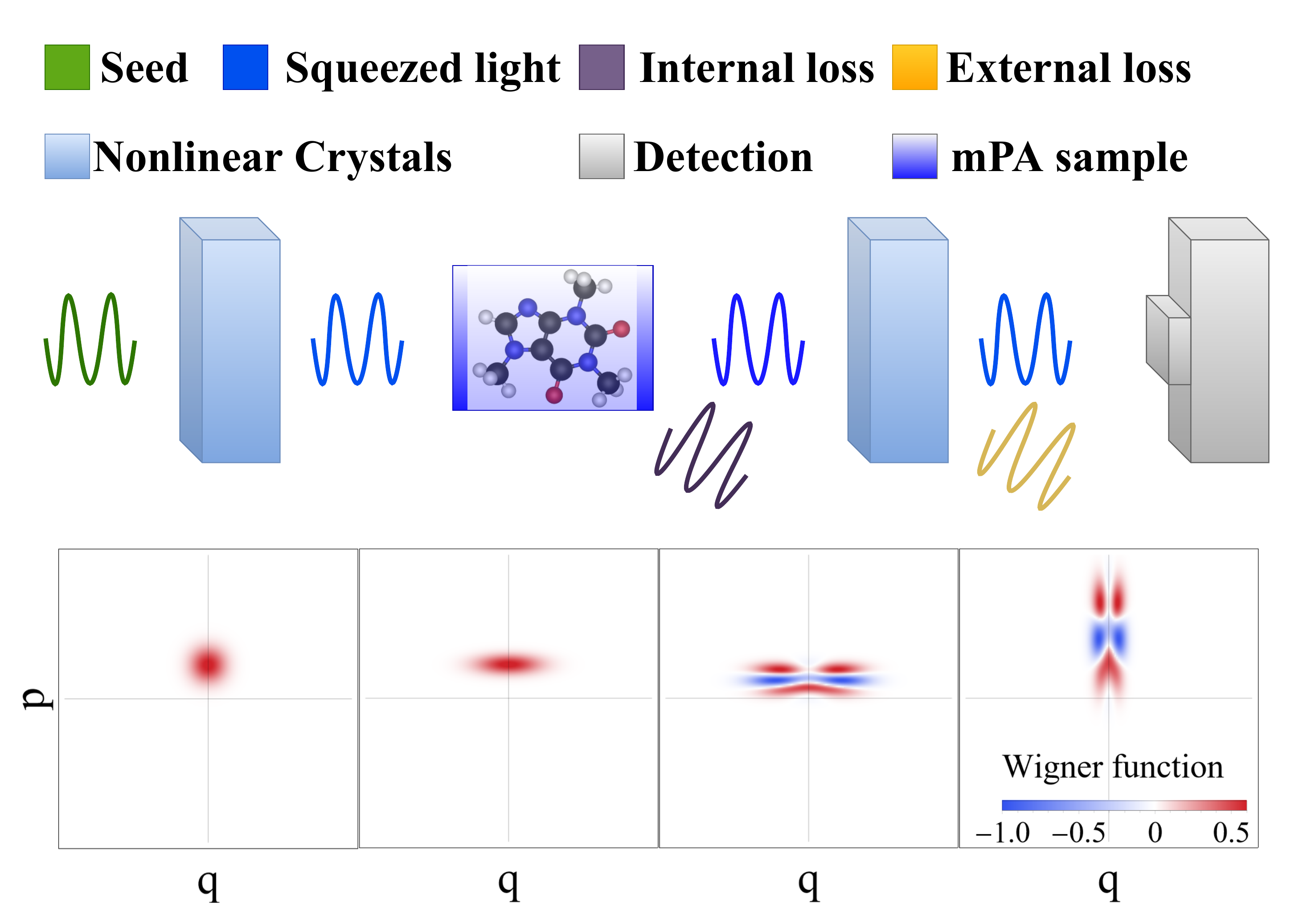}\llap{\parbox[b]{6.6in}{(a)\\\rule{0ex}{2in}}}
\includegraphics[width=0.23\textwidth]{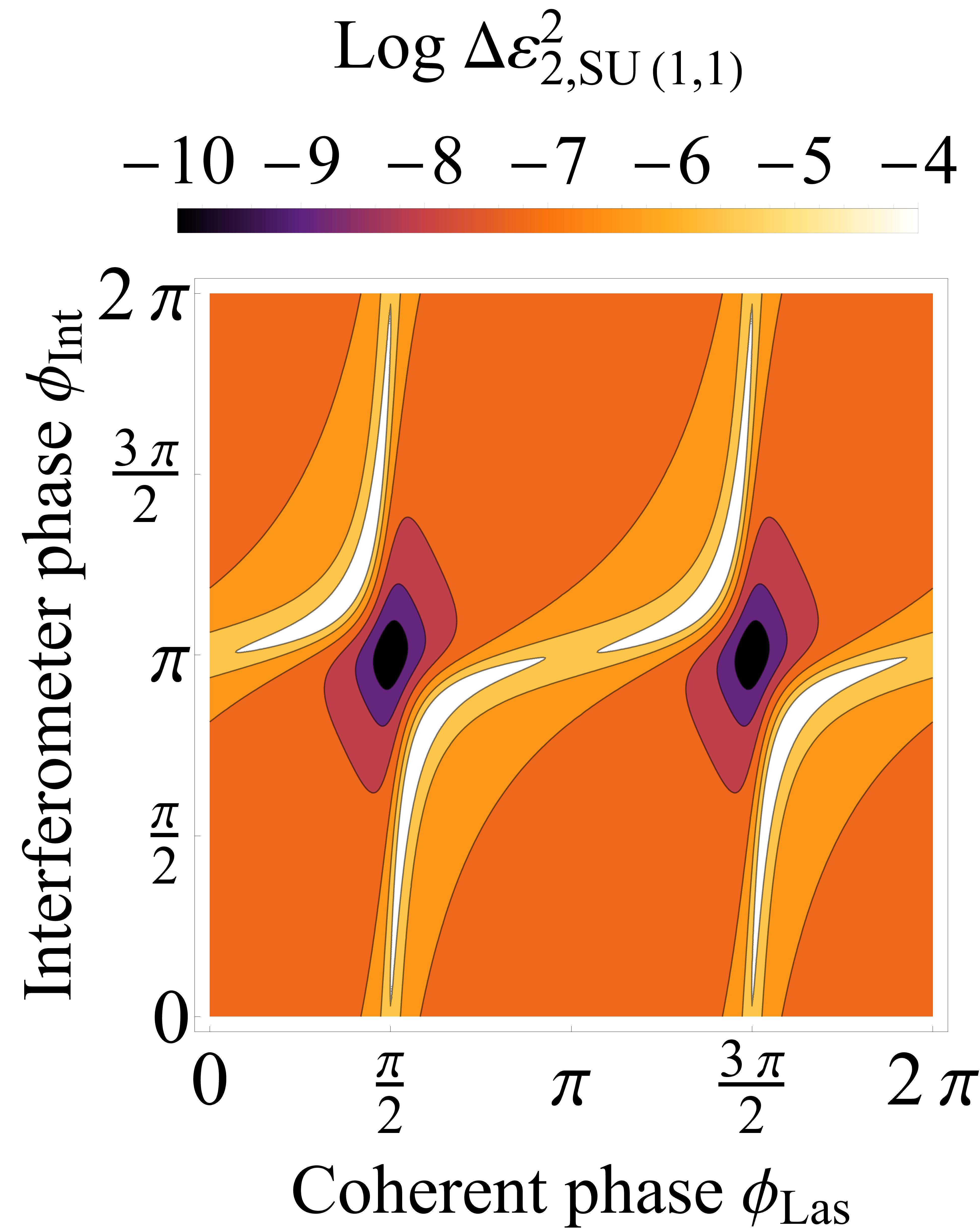}\llap{ \parbox[b]{3.1in}{(b)\\\rule{0ex}{2in}}}
\includegraphics[width=0.23\textwidth]{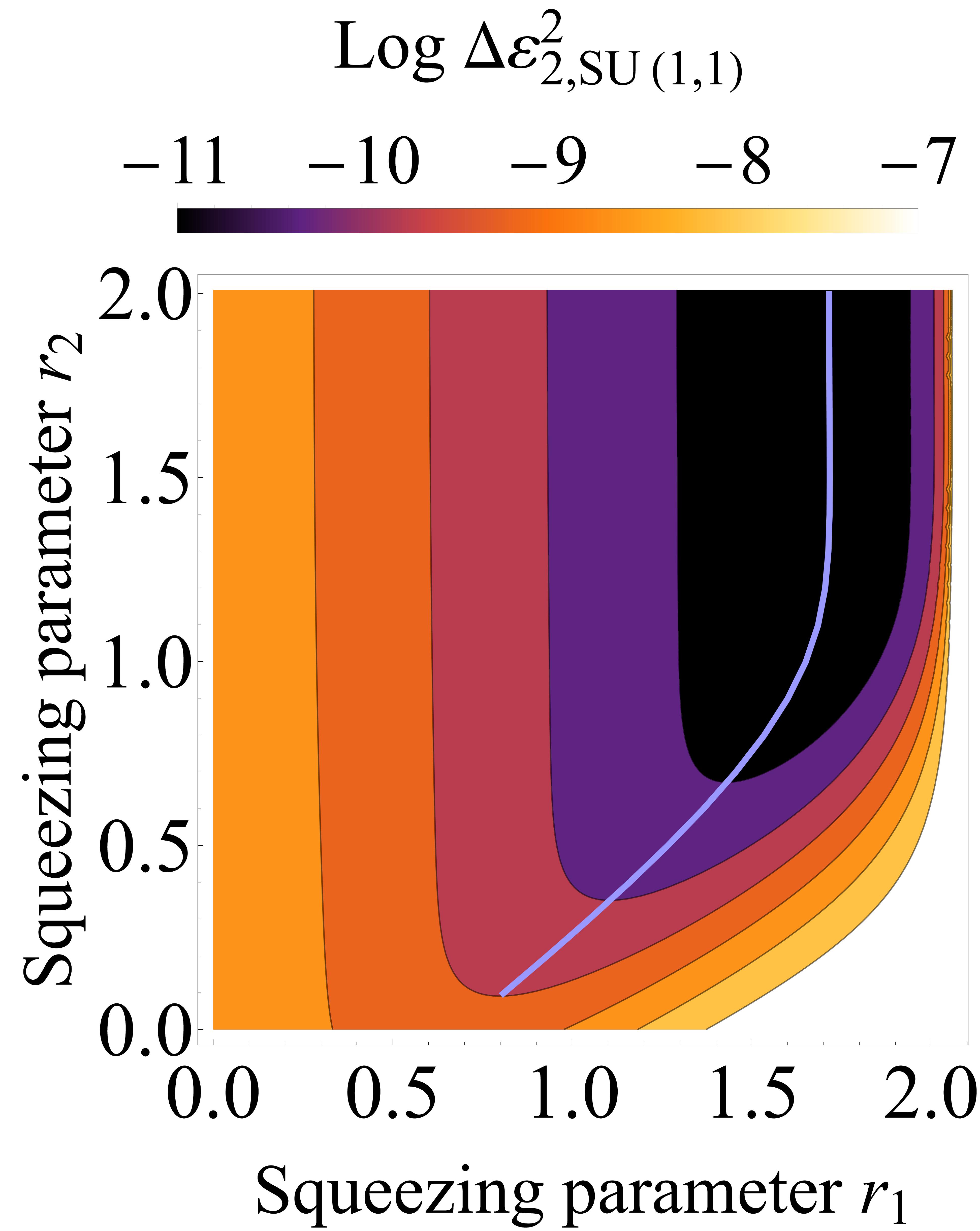}\llap{ \parbox[b]{3.1in}{(c)\\\rule{0ex}{2in}}}
\caption{ 
(a) The setup considered in this manuscript: A coherent seed pulse passes through a first degenerate OPA 
on the left which prepares a squeezed state. It is focused on an $m$-photon absorption sample, where a part of the light field is lost due to single photon scattering. 
The transmitted light field is then focused on a second degenerate OPA. The light field finally reaches the detection device on the right where imperfect detection gives rise to another loss source.
(b) Sensitivity for two-photon absorption, Eq.~(\ref{eq.sensitivity}) for $m = 2$, as a function of interferometer and laser phases is shown for a loss-free, ideal setup, $\eta_{In}=\eta_{Ex}=1$, as well as $r_{1}=0.939$, $r_{2}=1.447$, and $n_{S}=15$. 
(c) Variance $\Delta \varepsilon_{2,SU(1,1)}^2$, Eq.~(\ref{eq.sensitivity}), is shown as a function of squeezing parameters for $n_{S}=15$. The sensitivity becomes independent of the second squeezing parameter for a sufficiently large second squeezing parameter $r_2$. 
The light blue line maps optimal regimes of squeezing parameters.
}
\label{fig.setup}
\end{figure*}

One particularly promising platform for investigating the nonlinear interactions of bright quantum states of light are nonlinear SU(1,1)-interferometers~\cite{Chekhova16, Lawrie2019,Ou2020}. 
Introduced theoretically by Yurke already in the 1980's \cite{Yurke1986}, this technology has reached a level of maturity where various applications in quantum sensing seem feasible. 
It can be used, inter alia, for phase measurements~\cite{Manceau2017}, spectroscopy \cite{Michael19, Dorfman2020}, imaging \cite{Frascella19}, quantum state engineering \cite{Lemieux2016, Su2019}, and quantum information applications \cite{Shaked2018,Kalash2022}. The main difference between these nonlinear interferometers and their linear Mach-Zehnder counterparts is the replacement of the beam splitters with optical parametric amplifiers (OPA)~\cite{Ou2020}. In the single mode case considered in this manuscript, these OPAs squeeze (or anti-squeeze) a field quadrature of the light field propagating through the setup, and can thereby enhance resolution or suppress the impact of certain noise sources. 
This renders phase measurements robust against single photon losses that occur outside the interferometer~\cite{Marino12, Manceau17, Manceau2017, Liu18, Frascella19, Frascella20, Paterova20, Okamoto_2020,Du2022}. 
However, to the best of our knowledge, their use for absorption measurements has not been investigated to date.

In this Letter, we will show that by placing a mPA sample inside a nonlinear SU(1,1) interferometer~\cite{Chekhova16} and optimizing the interferometer, the resolution of a mPA signal can be enhanced significantly compared to mPA detection with classical means.
In particular, while the uncertainty $\Delta\varepsilon^2$ of $m$-photon absorption measurements with classical light scales as $\sim n_{S}^{-2m+1}$ for sufficiently large photon numbers ($n_{S}$) at the sample, we find that this scaling can be enhanced to $n_{S}^{-2m}$ in a nonlinear SU(1,1) interferometer, thus providing an enormous advantage in the mesoscopic photon number regime considered here.
We optimize theoretically the parameters of the interferometer such that this optimal sensitivity scaling can be realized, and we characterize the resulting light fields that are created inside the interferometer.
This analysis enables us to identify parameter regimes where nonlinear interferometers can detect mPA signals, but which are inaccessible with classical methods. 
In addition, we explore the effects of error sources (in literature referred to as internal and external losses \cite{Manceau17,Liu18,Frascella20}) on the mPA detection ability. Crucially, we find that so-called external losses do not degrade the sensitivity of the measurements, as they can be compensated in nonlinear interferometers. As is also the case in phase estimation~\cite{Ou2020}, however, so-called internal losses cannot be compensated, and reduce the optimal achievable precision scaling.
 
\emph{Setup and theory.---} The setup we consider in this paper is sketched in Fig.~\ref{fig.setup}(a):
a coherent seed field with frequency $\omega_0$ is injected into a degenerate OPA, where a pump pulse with frequency $2\omega_0$ triggers stimulated downconversion, and creates a squeezed coherent state that interacts with $m$-photon absorbing sample. 
We account for scattering losses in the optical system after mPA (internal losses), which we characterize by the loss rate $1-\eta_{In}$ [for single-photon losses occuring inside the sample, see the Supplementary Material (SM)]. The transmitted light field passes through the second OPA where it will be squeezed or anti-squeezed. Finally, the light field reaches the detector, where an intensity measurement is carried out. We account for imperfect photon detection with a second loss process (the so-called external losses), described by the loss rate $1-\eta_{Ex}$.
Taken together, the expectation value of the photon number measurement after the transmission through the described setup can be calculated as
\begin{align}
\langle \hat{n} \rangle &= \text{tr} \left\{ \hat{n} e^{\mathcal{L}_{loss \; 2}} e^{\mathcal{L}_{OPA \; 2} }  e^{\mathcal{L}_{loss \; 1}}  e^{ \mathcal{L}_{mPA} \varepsilon_m } e^{ \mathcal{L}_{OPA \; 1} }  \rho_0 \right\}. \label{eq.General}
\end{align}
Here, $\rho_0 = \mathcal{D} (\alpha) \vert 0 \rangle \langle 0 \vert \mathcal{D}^\dagger (\alpha)$ is the initial coherent seed state with amplitude $\alpha = \alpha e^{i \phi_{Las}}$, in which \textit{Las} stands for laser and we use $\alpha > 0$.  $\mathcal{L}_{OPA \; k}$ are superoperators describing the squeezing processes,
\begin{align}
e^{ \mathcal{L}_{OPA \; k} }  \rho &\equiv U_{OPA \; k} \rho U^{\dagger}_{OPA \; k}, \label{eq.U_OPA}
\end{align} 
where $U_{OPA \; k} = \exp (\zeta_k a^{\dagger 2}/2-\zeta_k^\ast a^2/2 )$ and $\zeta_k = r_k e^{i\phi_k}$. 
Without loss of generality, we set the induced phase of the first OPA as $\phi_{1}=0$ and rename the induced phase of the second OPA to $\phi_{2}=\phi_{Int}$ (where \textit{Int} stands for the interferometer).
$\mathcal{L}_{loss \; k}$ account for single-photon losses which we model as unbalanced beam splitters \cite{BarnettRadmore,Drummond,Gardiner,Manceau2017}, i.e.
\begin{align} \label{eq.L_loss}
e^{ \mathcal{L}_{loss \;  k}} \rho &= U_{loss \;k} \rho U^{\dagger}_{loss \;k},
\end{align}
in which $U_{loss \; k} = \exp \left(\tau_k \frac{a c^\dagger_k + c_k a^\dagger}{2} \right),$ with $\tau_k =  \arccos (\sqrt{\eta_k} )$, and $c_k$ is a photon annihilation operator in an auxiliary mode that remains in a vacuum state. 
Finally, the dynamics of the transmission of a quantum state of light through an $m$-photon absorbing medium is described by a Markovian Lindblad master equation for the photonic density matrix $\rho$ in a reference frame rotating at the frequency $\omega_0$~\cite{Agarwal1970,Zubairy1980}
\begin{align}
\frac{d}{dt} \rho &= \gamma_{mPA} \mathcal{L}_{mPA} \rho = \frac{\gamma_{mPA}}{2 m} \left( 2 a^m \rho a^{\dagger m} - a^{\dagger m} a^m \rho - \rho a^{\dagger m}  a^m\right). \label{eq.Lindblad}
\end{align}
The Lindblad operator is given by a correlated loss operator $L = a^m /\sqrt{m}$. We add the factor $m$ in the definition of the Lindblad operator for convenience to simplify expressions in our subsequent derivations. 
Given the time $t$ which the light field travels through the sample, we wish to estimate the absorbance $\varepsilon_m \equiv \gamma_{mPA} t$. It can be related to the corresponding $m$-photon absorption cross section (see SM). 
The precision of estimating $\varepsilon_m$ can be obtained by studying the uncertainty of $\varepsilon_m$ via error-propagation \cite{Toth2014,Niezgoda2019}. We focus on intensity measurements, hence the variance is given by
\begin{align}
\Delta\varepsilon_{m }^2 &= \frac{ \text{ Var } (\hat{n}) }{ \left| \frac{\partial \langle \hat{n} \rangle}{ \partial \varepsilon_m } \right|^2}. \label{eq.sensitivity}
\end{align}
Since nonlinear susceptibilities, and hence the corresponding multiphoton absorption cross sections, decline rapidly with $m$ (see SM), we concentrate on the weak absorption limit, where $\varepsilon_m \ll 1$. Consequently, we will approximate $e^{ \mathcal{L}_{mPA} \varepsilon_m } \simeq \mathbb{1}+\varepsilon_m \mathcal{L}_{mPA}$ which results into the transmitted density matrix being $\rho^\prime \simeq \rho + \varepsilon_m  (\partial\rho / \partial \varepsilon_m)$. 

\begin{figure*}
\centering
\includegraphics[width=0.24\textwidth]{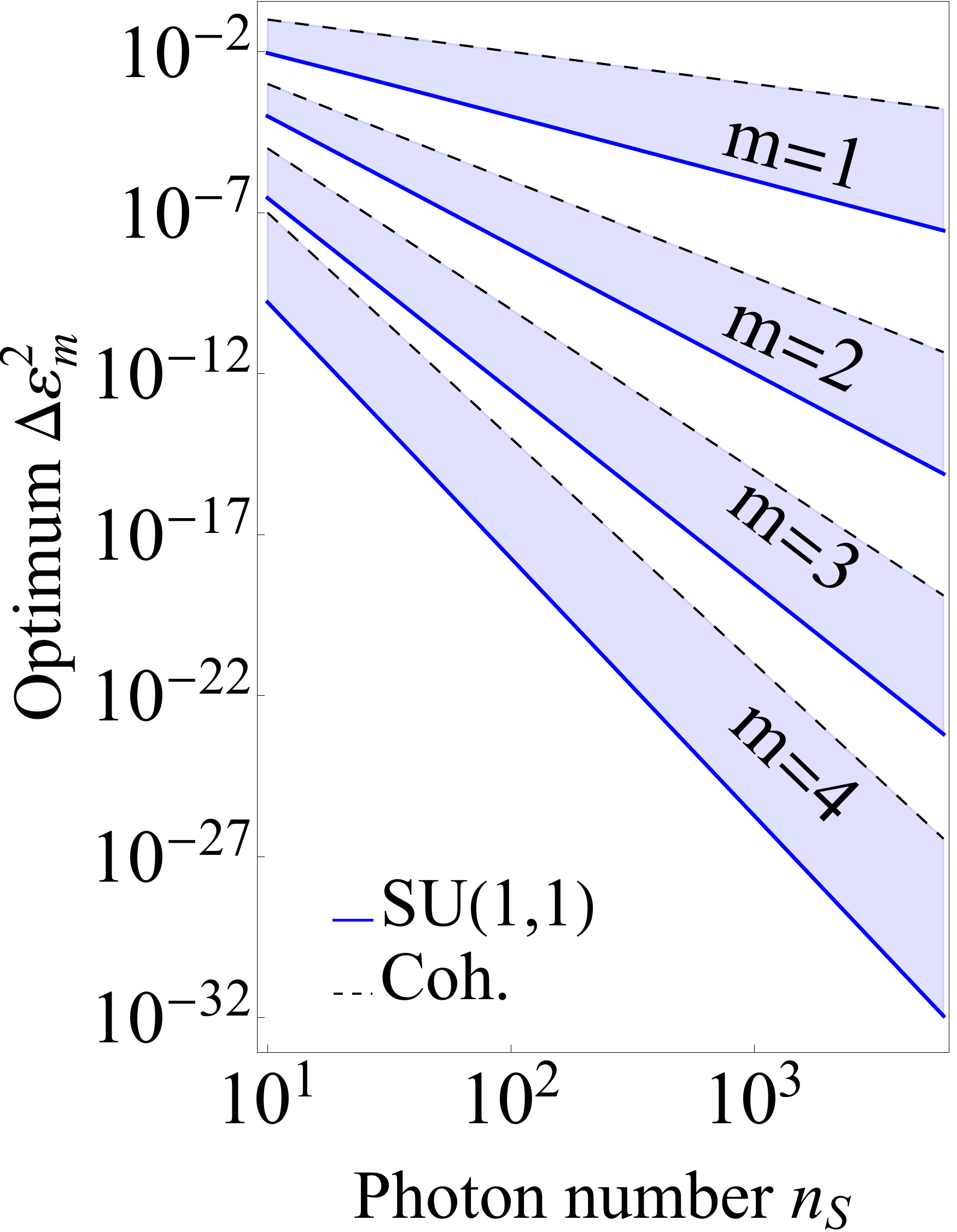}\llap{ \parbox[b]{3.2in}{(a)\\\rule{0ex}{2.2in}}}
\includegraphics[width=0.24\textwidth]{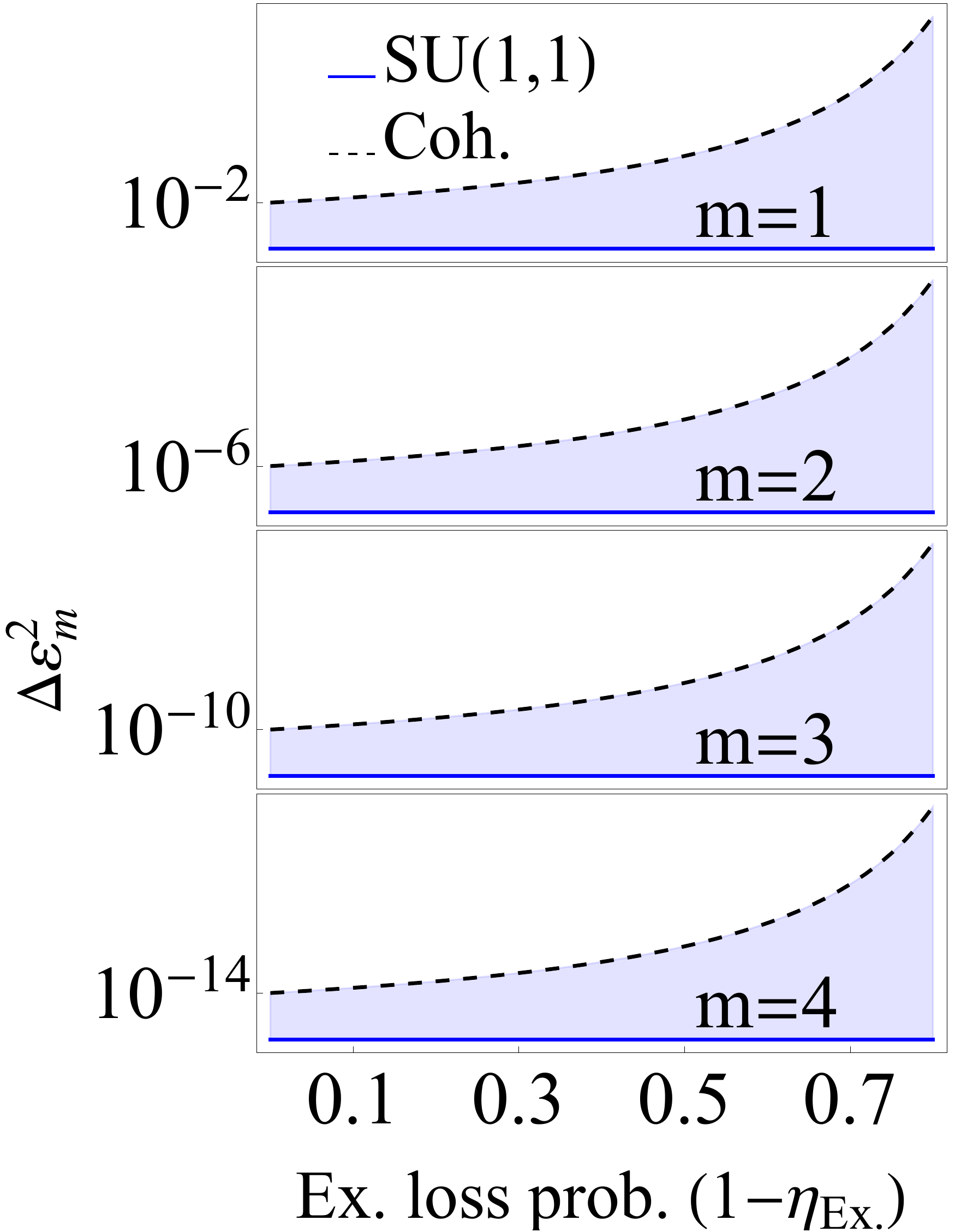}\llap{ \parbox[b]{3.2in}{(b)\\\rule{0ex}{2.2in}}}
\includegraphics[width=0.24\textwidth]{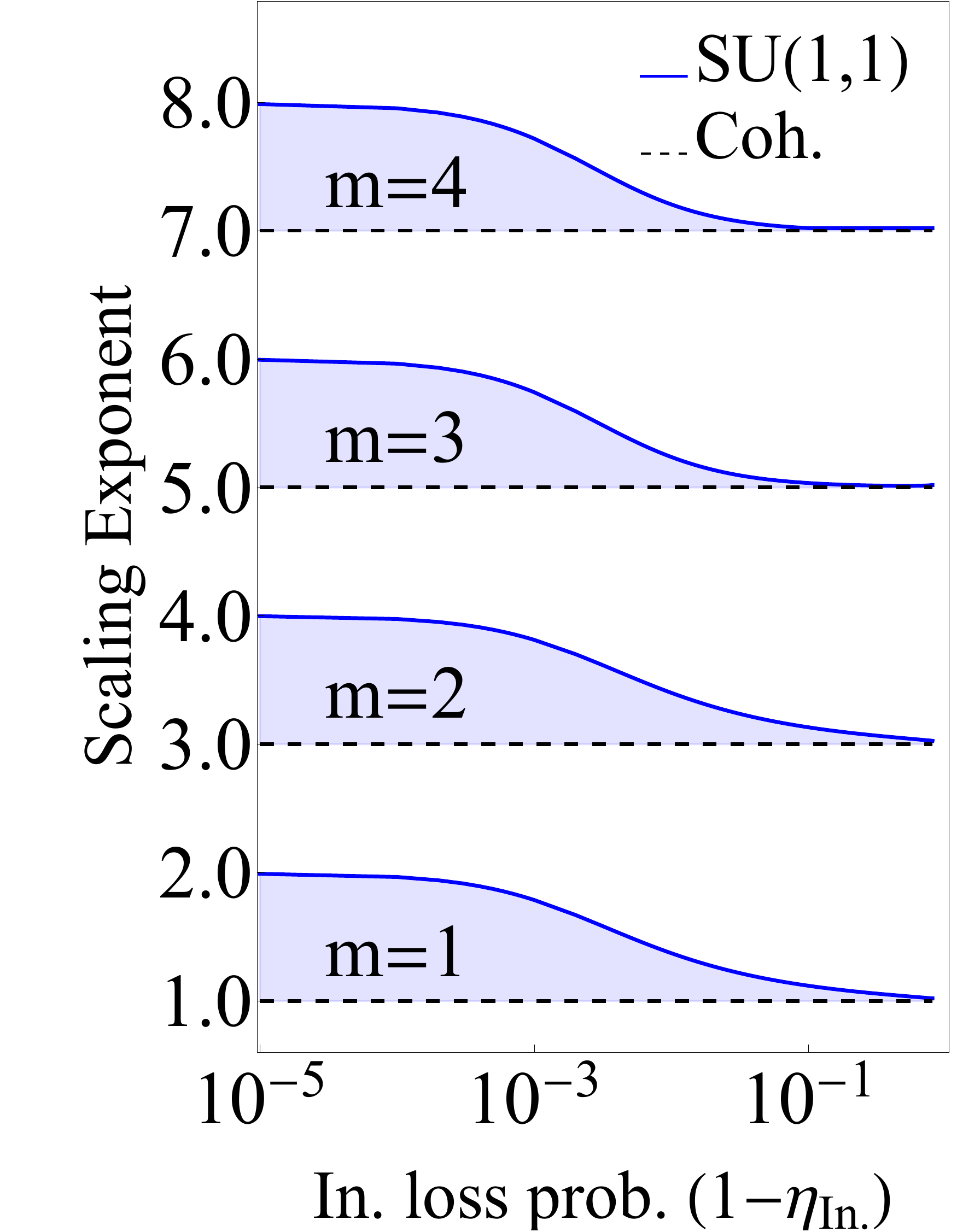}\llap{ \parbox[b]{3.2in}{(c)\\\rule{0ex}{2.2in}}}
\includegraphics[width=0.228\textwidth]{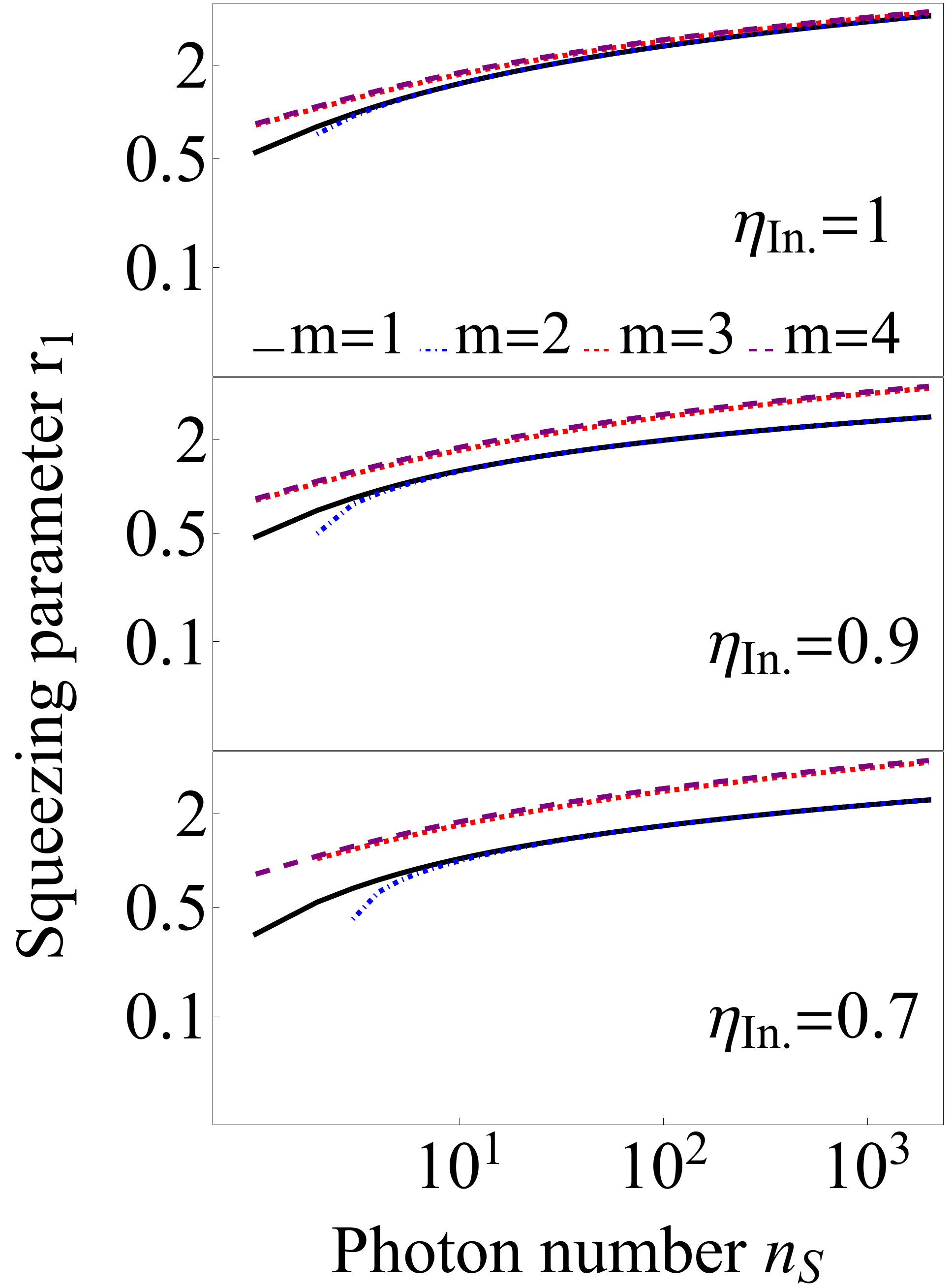}\llap{ \parbox[b]{3.1in}{(d)\\\rule{0ex}{2.2in}}}
\caption{ 
(a) Optimum $\Delta\varepsilon_{m }^2$ as a function of the photon number interacting with the sample, $n_S$. 
We observe that squeezing processes employed inside the SU(1,1) interferometer enhance the achievable optimum sensitivity compared to a classical interferometer. 
(b) Eq.~(\ref{eq.sensitivity}) is shown in the limit $r_2 \rightarrow \infty$ as a function of external losses for optimized SU(1,1) measurements (solid lines) and coherent states (dashed lines) in the case of one-, two-, three-, and four-photon absorption. 
By adjusting the squeezing parameters, the effects of external loss can be completely compensated in the SU(1,1) case.
(c) Scaling exponent $\gamma$ of the sensitivity (i.e. $\Delta\varepsilon_m^2 \sim n_{S}^{- \gamma}$) is shown as a function of the internal loss. 
SU(1,1) interferometric measurements shows superior precision scaling compared to coherent state measurements even in the presence of strong internal loss, e.g. $10\%$.
(d) The optimal squeezing parameter $r_1$ is shown as a function of photon number for three different levels of internal loss. 
Overall, we observe that the higher order nonlinear processes have better sensitivity indicating that the measurement of the cross section in samples with higher order nonlinear processes has better precision.}
\label{fig.results}
\end{figure*}

Let us first consider a conventional transmission measurement of mPA losses using a strong laser with $\langle \hat{n} \rangle \gg 1$. The precision can be found straightforwardly from Eqs.~(\ref{eq.General}) and (\ref{eq.sensitivity}) by setting $r_1 = r_2 = 0$. It evaluates to
\begin{align}
\Delta\varepsilon_{m, (coh)}^{2} &= \frac{1}{ \eta_{In}\eta_{Ex}}\frac{1}{n_{S}^{2m-1}}, \label{eq.scalingDegrCoh}
\end{align}
where $n_S$ (the number of photons at the sample) in Eq.~(\ref{eq.scalingDegrCoh}) is simply $n_S = \alpha^2$. The precision scales as the $2m-1$-th power of the mean photon number, and it is reduced by both internal and external losses.

\emph{Interferometric enhancement.---}We now want to understand how the SU(1,1) interferometer can enhance the resolution of mPA measurements compared to Eq.~(\ref{eq.scalingDegrCoh}). To this end, we first establish conditions for optimizing the sensitivity, i.e. for minimizing the variance~(\ref{eq.sensitivity}). 
Since the photon number in the interferometer can vary strongly with, e.g., the different phases $\phi_i$, we fix the number of photons at the sample ($n_{S}$) in the optimization procedure. This requirement is inspired by the central objective in quantum-enhanced sensing applications to reduce photodamage at the sample. Instead, we allow for a varying number of photons at the detector level, where photodamage is not problematic.
As a consequence, the first squeezing parameter is bounded within $0 \leqslant r_{1} \leqslant \text{arcsinh} \sqrt{n_{S}}$, and the coherent amplitude $\alpha$ is adjusted to keep the photon number at the sample fixed resulting in
$
n_{S} = \alpha^2 \bigg[\sinh (2 r_{1}) \cos (2 \phi_{Las})+\cosh (2 r_{1})\bigg]+\frac{1}{2} \bigg[ \cosh (2 r_{1})-1 \bigg]
$.
Thus, $r_1 = \text{arcsinh} \sqrt{n_{S}}$ will correspond to a squeezed vacuum (and also $\alpha = 0$), while $r_1 = 0$ corresponds to a coherent state with $\alpha = \sqrt{n_S}$ propagating through the sample.

In Fig.~\ref{fig.setup}(b), we present the parameter optimization for 
$m = 2$ at a fixed photon number $n_S = 15$ showing that we obtain the greatest precision when the phases are fixed at $\phi_{Las} =\pi/2, 3\pi /2, \ldots$ and $\phi_{Int}=\pi$. 
This result does not change for $m = 1,\ldots 4$ \cite{Fnote} 
and does not depend on the photon number $n_S$. 
It can be interpreted as illustrated in the small panels in Fig.~\ref{fig.setup}(a): The first condition implies that the first OPA generates an amplitude-squeezed state (2nd panel from left). We recently showed that such a state is well suited to detect two-photon losses \cite{Panahiyan2022}, and this remains true for general $m$-photon absorption. The second condition above, $\phi_{Int}=\pi$, implies that the second OPA anti-squeezes the initially squeezed quadrature. As illustrated in the third and fourth small panel in Fig.~\ref{fig.setup}(a), this is because the mPA information is most strongly imprinted on the squeezed quadrature. By anti-squeezing it, the second OPA thus magnifies the changes brought about by the mPA process on the light field.

The optimal choice of phases has to be accompanied by a suitable degree of squeezing. This is depicted in Fig.~\ref{fig.setup}(c), where we minimize Eq.~(\ref{eq.sensitivity}) as a function of both squeezing parameters $r_1$ and $r_2$. For any fixed $r_2$, there is a minimum as a function of $r_1$. This means that the optimal sensitivity is generated by an amplitude-squeezed state, where the amount of squeezing has to be adjusted for the mean photon number $n_S$, as we will discuss later in detail. 
In addition, the optimal squeezing $r_1$ increases with $r_2$ and then saturates, as does the achievable precision at fixed $n_S$. This demonstrates that the interferometer can enhance the precision of mPA detection. 

This finding motivates us to investigate the behaviour of Eq.~(\ref{eq.sensitivity}) in the limit $r_2 \rightarrow \infty$, i.e. in a limit where the second OPA is much stronger than the first.
If we optimize $r_1$ at every $n_S$ (for details, see SM), Eq.~(\ref{eq.sensitivity}) scales in the absence of any losses as 
\begin{align}
\Delta\varepsilon_{m, SU(1,1)}^2 &\propto \frac{1}{n_{S}^{2m}} \label{eq.scaling}
\end{align}
for sufficiently large $n_S$.  
This scaling is contrasted with the classical case~(\ref{eq.scalingDegrCoh}) in Fig.~\ref{fig.results}(a) for $m = 1...,4$. Thus, the optimal choice of parameters enables us to increase the scaling of the sensitivity with respect to the mean photon number by a factor one. This is the central result of our paper. 

\emph{Impact of losses.---} We next investigate how the quantum-enhanced sensitivity scaling is affected by the inevitable photon losses in a realistic experiment. 
We first note that in the limit $r_2 \rightarrow \infty$ considered before, Eq.~(\ref{eq.sensitivity}) becomes independent of external losses (see SM). 
This is again consistent with our earlier results~\cite{Panahiyan2022}, where it was shown that single-photon losses do not affect the sensitivity of two-photon absorption measurements with squeezed states.
Hence, external losses can always be compensated in an SU(1,1) interferometer~\cite{Manceau17, Lemieux2016}. 
This behaviour contrasts with the classical situation in Eq.~(\ref{eq.scalingDegrCoh}), where the resolution is degraded from external losses. This behaviour is shown in Fig.~\ref{fig.results}(b).

The same is not true of internal losses, unfortunately. If these losses destroy squeezing in the light field before it reaches the second OPA, the quantum advantage we obtained in Eq.~(\ref{eq.scaling}) is lost. This behaviour is shown in Fig.~\ref{fig.results}(c), where we extract the scaling behaviour of $\Delta\varepsilon_m^2$ (i.e. we determine the scaling $\Delta\varepsilon_m^2 \sim n_S^{-\gamma}$ for $n_S > 100$) as a function of these internal losses. At small internal losses ($1 - \eta_{In} \lesssim 10^{-3}$), the scaling is not affected and we find $\gamma \sim 2 m$. The scaling exponent then decreases gradually, and approaches the coherent limit $\gamma \sim 2m -1$ when the internal losses become very large ($1 - \eta_{In}\sim 1$).  
It should be noted that although the scaling behavior of the coherent case is not affected by internal loss, the optimum sensitivity would be affected similarly to the external case [see Eq.~(\ref{eq.scalingDegrCoh})]. 
There is a quantum enhancement (albeit a small one) even for strong internal losses of, say, $10\%$. 
Remarkably, overall our findings appear analogous to linear phase estimation applications, where internal losses destroy the sought-after Heisenberg scaling, while external losses can be compensated~\cite{Marino12, Frascella19, Frascella20, Manceau17, Manceau2017, Liu18, Paterova20}.

\emph{Optimal squeezing parameter.---} We finally turn to the discussion of the optimal squeezed state in the interferometer. 
The observed enhancement is a consequence of the optimal squeezing/anti-squeezing operations performed by the two OPAs. While the second squeezing parameter should be chosen as large as possible, the first squeezing parameter is determined by three factors: the number of photons $n_S$, the internal loss $1-\eta_{In}$, and degree of the nonlinearity $m$. 
We show its variation with $n_S$ in Fig.~\ref{fig.results}(d). 
In an ideal interferometer, $\eta_{In} = 1$, the first squeezing generates an amplitude-squeezed state
[see Fig.~\ref{fig.setup}(a)] where the standard deviation of the anti-squeezed quadrature is as large as the expectation value of the squeezed one, i.e. $\langle q^2 \rangle^{1/2} = \langle p \rangle$. As we show in the SM, this condition gives rise to the enhanced scaling in Eq.~(\ref{eq.scaling}), by keeping the variance in the numerator of Eq.~(\ref{eq.sensitivity}) constant, while admitting an optimal scaling of the denominator. 
This optimal degree of squeezing decreases with the internal loss in the case of one- and two-photon absorption but remains almost constant for larger nonlinearities. Thus, the condition $\langle q^2 \rangle^{1/2} = \langle p \rangle$ appears to describe an almost universally optimal state for detecting multiphoton absorption. 

\emph{Conclusion.---} We found that SU(1,1) interferometers can enhance the precision of multi-photon absorption measurements compared to classical approaches, thus reducing the necessary photon flux to observe a signal. 
At photon fluxes with mean photo number $\langle\hat{n}\rangle > 1$, 
the precision of an optimally tuned SU(1,1) interferometer scales as $\sim n_{S}^{-2m}$. In contrast, a classical measurement only realizes a scaling $\sim n_{S}^{-2m+1}$. A measurement with the optimized ideal SU(1,1) interferometer thus outperforms its classical counterpart by a factor one. 
Even in the presence of competing losses, optimized quantum states always outperform classical measurement strategies. 
We note that this performance is also superior to transmission measurements with squeezed light (see ref.~\cite{Panahiyan2022}).
While we concentrated on multiphoton absorption, our work suggests interesting quantum enhancements for other nonlinear spectroscopic processes such as Raman signals~\cite{Michael19}. 
The generalization to multimode interferometers will be another important avenue to explore, as it will enable the exploitation of entanglement to create further metrological and spectroscopic advantages in nonlinear interferometry~\cite{Asban2021,Asban2021b, Dorfman2021b, Asban2022}. 
Finally, our work establishes a quantum advantage for high-gain squeezed light that opens new applications in quantum imaging~\cite{Moreau2019, Lemos2022_review} in a nonlinear intensity regime.

\begin{acknowledgments} 
 
S. P. and F. S. acknowledge support from the Cluster of Excellence 'Advanced Imaging of Matter' of the Deutsche Forschungsgemeinschaft (DFG) - EXC 2056 - project ID 390715994. C. S. M. acknowledges that the project that gave rise to these results received the support of a fellowship from la Caixa Foundation (ID 100010434) and from the European Union Horizon 2020 research and innovation program under the Marie Skodowska-Curie Grant Agreement No. 847648, with fellowship code LCF/BQ/PI20/11760026, and financial support from the Proyecto Sin\'ergico CAM 2020 Y2020/TCS-6545 (NanoQuCo-CM). 

\end{acknowledgments}

\bibliography{bibliography_photons}

\begin{thebibliography}{72}
\bibitem [{Fno()}]{Fnote}%
  \BibitemOpen
  \href@noop {} {}\bibinfo {organization} {Even though we cannot completely
  rule out the possibility of distinct parameters for larger $m$ which give
  rise to even better resolution, we prove the improved scaling,
  Eq.~(\ref{eq.scaling}), for any $m$ in the appendix.}\BibitemShut {Stop}%
\end{thebibliography}

\newpage

\begin{widetext}
\renewcommand{\theequation}{A.\arabic{equation}}
\setcounter{equation}{0}

\section{Supplementary material}

\section{Lindbladian approach to m-photon absorption}

In this section, by drawing heavily on Refs. \cite{Agarwal1970,Zubairy1980}, we provide details of derivation of Lindbladian for mPA given in Eq. \eqref{eq.Lindblad}. We assume that our sample consists of an ensemble of the systems that have two levels, i.e. $\vert 1 \rangle$ and $\vert 2 \rangle$. The transition takes place from state $\vert 1 \rangle$ to $\vert 2 \rangle$ by the sample absorbing $m$ photons. We consider this $m$-photon transition resonant, hence $\omega_{12}=m \omega_{0}$. The interaction Hamiltonian is given by 
\begin{align}
H_{I}=\sum_{i} (\xi \sigma_{2 i}^{\dagger} \sigma_{1 i}E^{+ m}(\Vec{r}_{i})+H.c.),
\end{align}
in which $\xi $ is the matrix element for mPA, $\sigma_{1 i}^{\dagger}$ and $\sigma_{2 i}^{\dagger}$ are creation operators for $i$th two-level system in states of $\vert 1 \rangle$ and $\vert 2 \rangle$, respectively. The positive-frequency part of the electric field at the $i$th two-level system 
is given by
\begin{align}
E^{+}(\Vec{r}_{i})=- i \sqrt{\frac{\hbar \omega}{2 \epsilon_{0}}} u(\Vec{r}_{i})a,
\end{align}
where $u(\Vec{r}_{i})$ are the normalized mode eigenfunctions. The equation of motion for the density operator of combined sample-photon field is
\begin{align}
i \hbar \frac{\partial \rho_{T}(t)}{\partial t}=\left[\hat{H}_{I}^{\prime}, \rho_{T}(t)   \right],
\end{align}
in which $\hat{H}_{I}^{\prime}$ is the Hamiltonian in interaction picture. At $t=0$, the two-level systems of the sample are decoupled from photon field and
\begin{align}
\rho_{T}(0)=\rho(0) \bigotimes \prod_{i} \rho(0)_{i},
\end{align}
where $\rho(0)_{i}$ is the thermal-equilibrium density operator for the $i$th two-level system. The density operator of the photon-field at time $t$ can be obtained by taking trace over two-level systems of the total density operator $\rho(t)=\text{Tr}_{i}{\rho_{T}(t)}$. We can obtain the master equation by using the standard perturbation techniques based on Born-Markov approximations 
\begin{align}
 \frac{\partial \rho(t)}{\partial t}=\frac{\kappa_{1} \gamma_{mPA}}{2m}\left(\left[a^m \rho(t), a^{\dagger m}  \right]+\left[a^m, \rho(t) a^{\dagger m}  \right] \right)+ 
\frac{\kappa_{2} \gamma_{mPA}}{2m}\left(\left[a^{\dagger m} \rho(t), a^m  \right]+\left[a^{\dagger m} , \rho(t) a^m  \right] \right), \label{eq.totalLin}
\end{align}
in which $\kappa_{1}$ ($\kappa_{2}$) is the thermal population of the two-level systems and and $\gamma_{mPA}$ is given by \cite{Agarwal1970,Zubairy1980}
\begin{align}
\gamma_{mPA}=2m (\frac{\hbar}{2 \epsilon_{0}})^{m-2}(2 \pi)^2 \omega^m |\zeta|^2 g(m \omega)
\int d^3r N(\Vec{r}) |u(\Vec{r})|^{2m}, 
\end{align}
in which $g(m \omega)$ is the line-shape function and $N(\Vec{r})$ is the density of two-level systems at the position $\Vec{r}$ and integration takes place over volume of the medium. The first term in Eq. \eqref{eq.totalLin} describes the absorption process while the second term accounts for the emission process. By considering the sample at zero temperature, ( $\kappa_{2}=0$ and $\kappa_{1}=1$),  \eqref{eq.totalLin} reduces to the Lindbladian in \eqref{eq.Lindblad}.

\section{Heisenberg picture calculations} \label{sec:Heisenberg}

Here we describe how the optical signals and the measurement precision can be calculated straightforwardly in the Heisenberg picture. 
The squeezing operations in the two OPA's can be described by the following transformation on the photon annihilation operator~\cite{BarnettRadmore}: 
\begin{align}
a \rightarrow a' = \cosh (r_k) a + \sinh (r_k) e^{i \phi_k} a^\dagger. \label{eq.U_OPAk}
\end{align}

The imperfections in the experimental setups, i.e. single-photon losses, were modeled by beam splitter transformations. The beam splitter transformation approach to single-photon losses is well established for the description of imperfect photon detection or losses in the optical setup \cite{Drummond,Manceau2017,Manceau17}. To describe scattering losses in the mPA medium rigorously, however, it would be necessary to include them in the evolution equation~(\ref{eq.Lindblad}). 
This is not an issue for our analysis, as we evaluate the measurement precision in the limit $\varepsilon_m \ll 1$, but should be considered at finite absorbances.  
Modeling single-photon losses as unbalanced beam splitters, we obtain the input-output relation from Eq.~(\ref{eq.L_loss}) as~\cite{Drummond,Manceau2017,Manceau17}
\begin{align}
a^{(out)} &= \sqrt{\eta_k} \; a + \sqrt{1 - \eta_k} \; c_k, \label{eq.lin-loss}
\end{align}
where $c_k$ is the vacuum (empty port) photon annihilation operator. Using transformations in Eqs.~\eqref{eq.U_OPAk} and \eqref{eq.lin-loss}, we can find the expectation value of photon number in the following form: 
\begin{align}
\frac{\partial \langle \hat{n} \rangle}{ \partial \varepsilon }  \bigg|_{\varepsilon = 0}
&= \text{tr} \left\{ \hat{n}_{f}  \rho_0 \right\}, \label{eq.derivativeNew}
\end{align}
in which 
\begin{align}
&\hat{n}_{f}=  
U^{\dagger}_{loss \;2}U^{\dagger}_{OPA \;2}U^{\dagger}_{loss \;1}\mathcal{L}_{mPA}^{\prime \; adj}[U^{\dagger}_{OPA \;1} \hat{n} U_{OPA \;1}] U_{loss \;1} U_{OPA \;2} U_{loss \;2}, \label{newN}
\end{align}
where for a general operator $X$, $\mathcal{L}_{mPA}^{\prime \; adj}$ performs the following operation: 
\begin{align}
\mathcal{L}_{mPA}^{\prime \; adj}[X]=\frac{1}{2 m} \left( 2 a^{\prime \; \dagger m} X a^{\prime m}- a^{\prime \; \dagger m} a^{\prime m}  X -  X a^{\prime \; \dagger m} a^{\prime m} \right), \label{eq.LindbladNew}
\end{align}
and the corresponding operators in Eq. \eqref{newN} are given by the transformations in Eqs. \eqref{eq.laser}-\eqref{eq.trafo-etaEx}. The result for expectation value \eqref{eq.derivativeNew} is given in the Schr\"odinger picture. It is more convenient to perform the calculations in Heisenberg picture. To do so, we introduce the following transformations in Heisenberg picture for photons operators to evaluate Eq. \eqref{eq.derivativeNew}: 

\begin{enumerate}[label=\arabic*)]
\item We apply the displacement transformation which gives us 
\begin{align}
a \rightarrow a+\alpha e^{i \phi_{Las}} \label{eq.laser},
\end{align}
\item We apply the squeezing transformation (Eq. \eqref{eq.U_OPAk}) to account the first squeezing operation
\begin{align}
a \rightarrow a' = \cosh (r_1) a + \sinh (r_1) a^\dagger \label{eq.OPA1},
\end{align}
in which, we have set $\phi_{1}=0$ without loss of generality.
\item Next, we apply the adjoint Lindbladian~(\ref{eq.LindbladNew}) to $a'$.
\begin{align}
a' \rightarrow a'' &= a' + \varepsilon_{m}\mathcal{L}^{adj}_{mPA} [a'] = a' - \frac{\varepsilon_{m}}{2}a'^{\dagger \ m-1}a'^m. \label{eq.Lindblad-trafo}
\end{align}
\item For the internal loss, we perform the following transformation by using Eq. \eqref{eq.lin-loss}: 
\begin{align}
a'' \rightarrow a_2'' = \sqrt{\eta_{In}} a'' + \sqrt{1 - \eta_{In}} c_1, \label{eq.trafo-etaIn}
\end{align}
where for full justification on modelling single-photon loss by beam splitter, please see appendix of Refs. \cite{Panahiyan2022} or Refs. \cite{BarnettRadmore,Drummond,Gardiner}.
\item We perform the squeezing operation with respect to the second OPA, which is described by the transformation $U^\dagger_{OPA \; 2}$ and Eq. \eqref{eq.U_OPAk}:
\begin{align}
a_2'' \rightarrow a''' = \cosh (r_2) a'' + e^{i \phi_{Int}} \sinh (r_2) a''^{\dagger} \label{eq.squeezing-trafo}.
\end{align}
\item Finally, to account for the external loss and by considering \eqref{eq.lin-loss}, we apply the following transformation: 
\begin{align}
a''' \rightarrow a_2''' = \sqrt{\eta_{Ex}} a''' + \sqrt{1 - \eta_{Ex}} c_2. \label{eq.trafo-etaEx}
\end{align}
\end{enumerate}

Finally, the expectation value of the transformed photon number operator $\hat{n}'' = a^{'' \dagger} a''$ is taken with respect to the vacuum state and the linear terms in $\varepsilon_m$ are collected. 

\begin{figure*}
\centering
\includegraphics[width=0.24\textwidth]{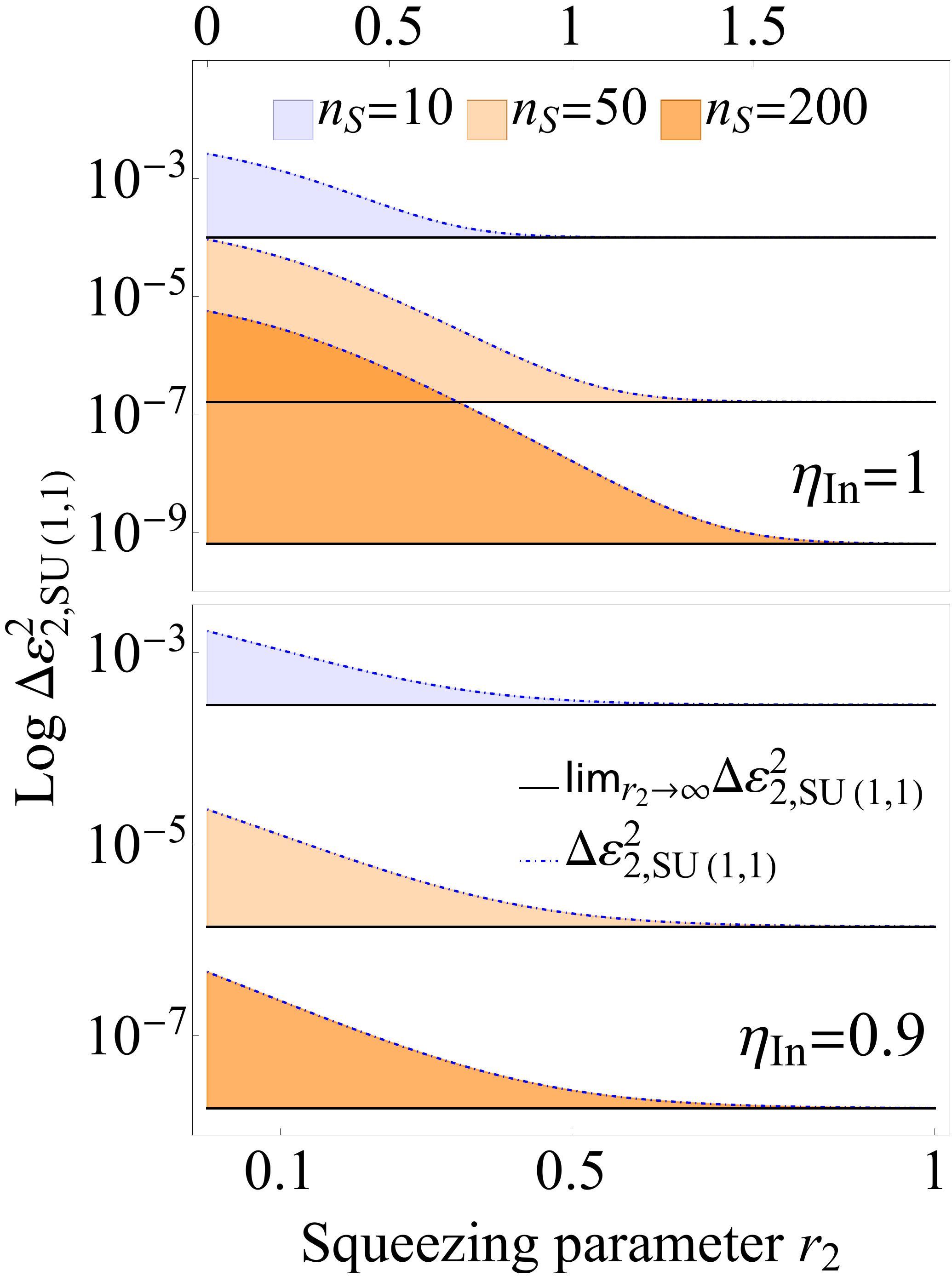}\llap{ \parbox[b]{3.2in}{(a)\\\rule{0ex}{2.3in}}}
\includegraphics[width=0.32\textwidth]{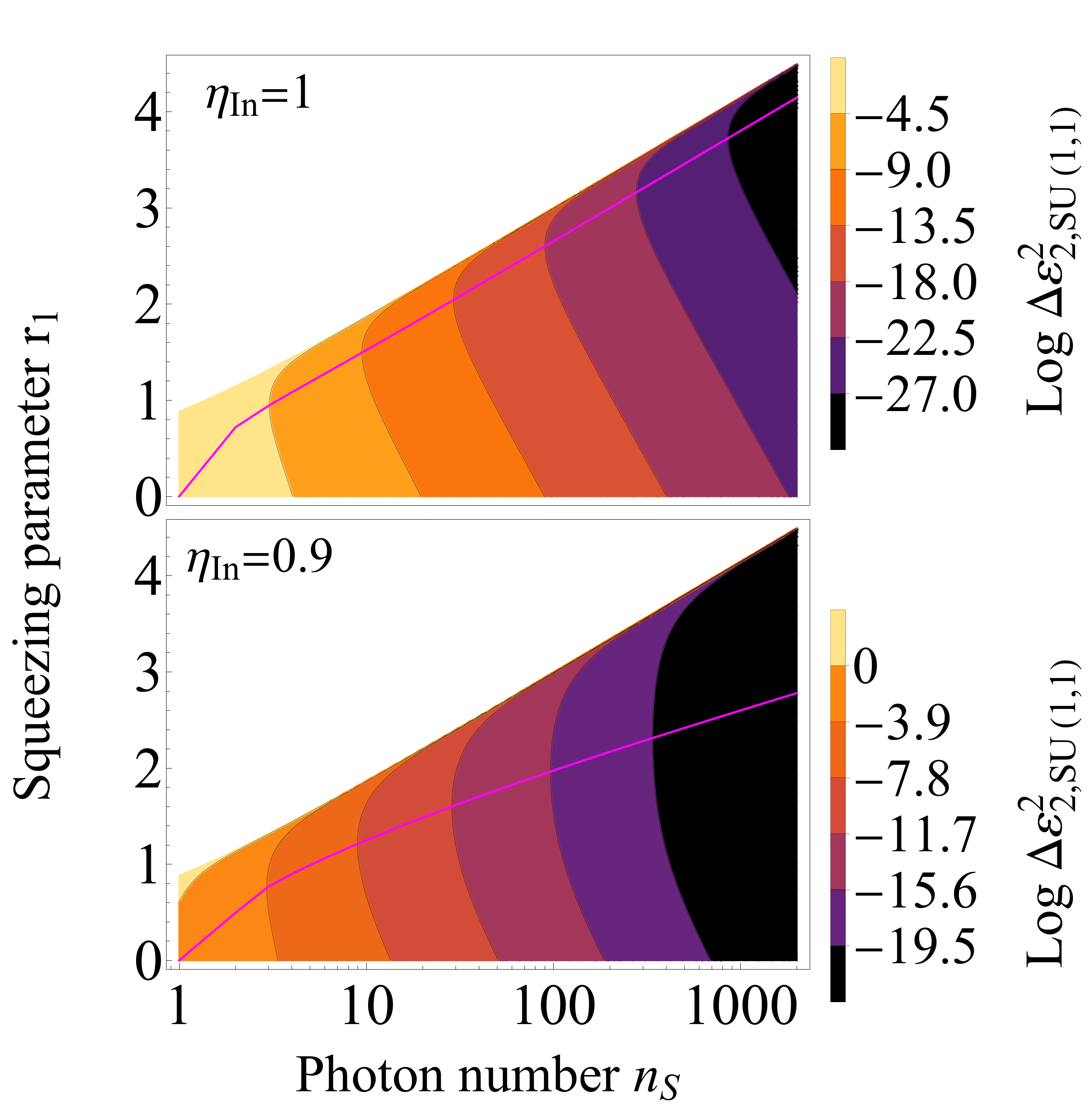}\llap{ \parbox[b]{4.3in}{(b)\\\rule{0ex}{2.3in}}}
\includegraphics[width=0.245\textwidth]{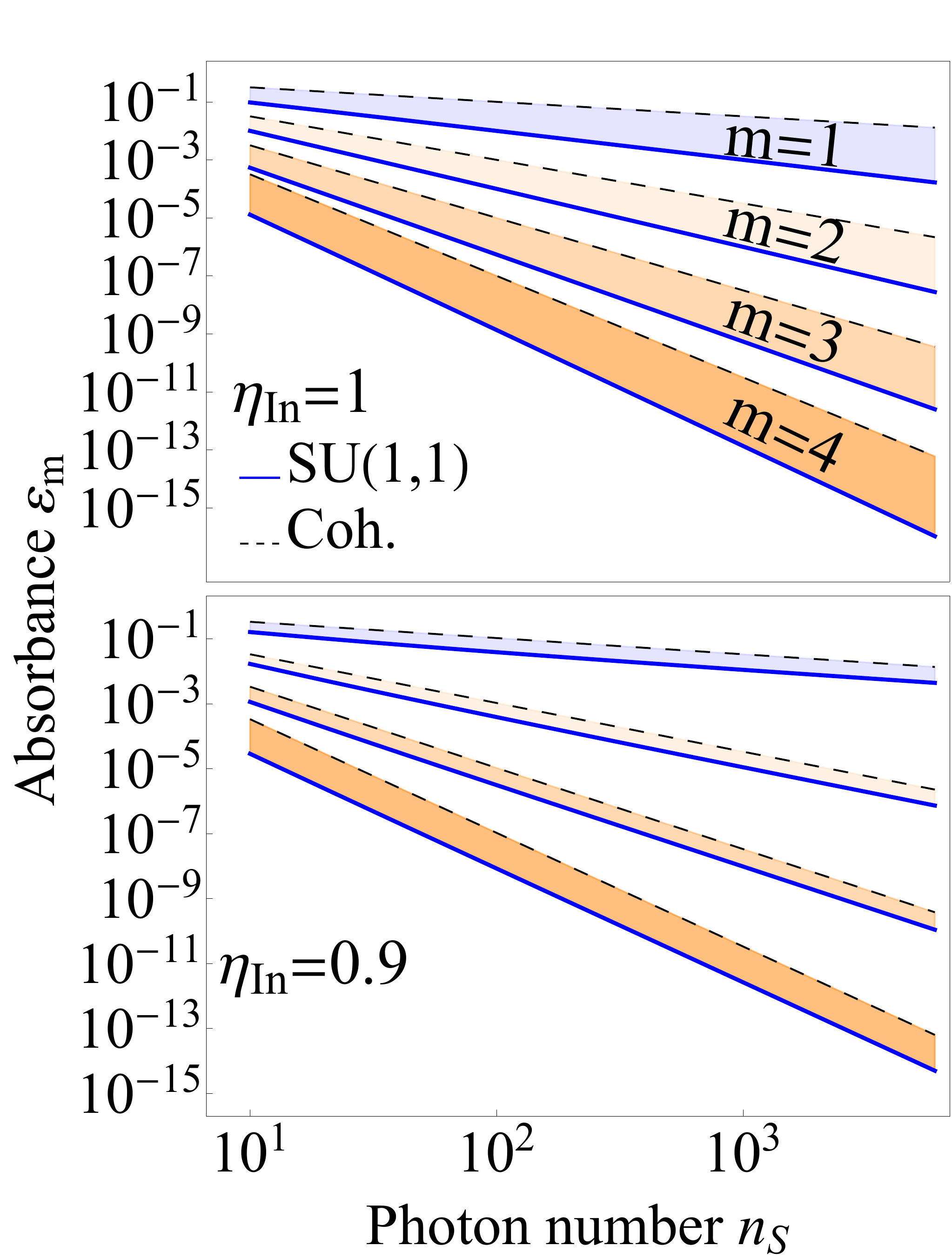}\llap{ \parbox[b]{3.2in}{(c)\\\rule{0ex}{2.3in}}}
\caption{ 
(a) Variance $\Delta\varepsilon^2_{2}$ for two-photon absorption as a function of the second squeezing parameter in cases of $\Delta\varepsilon_{2, SU(1,1)}^2$ (dotted-dashed lines) and $\Delta\varepsilon_{2, SU(1,1)}^2 \bigg|_{r_{2} \rightarrow \infty}$ (solid lines) for different photon numbers at the sample, $n_S$. We observe that for sufficiently large second squeezing parameter, $\Delta\varepsilon_{2, SU(1,1)}^2$ converges to $\Delta\varepsilon_{2, SU(1,1)}^2 \bigg|_{r_{2} \rightarrow \infty}$. 
(b) Logarithm of $\Delta\varepsilon_{2, SU(1,1)}^2 \bigg|_{r_{2} \rightarrow \infty}$ as a function of the first squeezing parameter and photon numbers. The magenta lines map the optimum regimes of first squeezing parameter with their corresponding photon numbers. 
(c) The smallest absorbance that can be detected for a given photon number $n_S^{min}$, is shown for optimal regime of sensitivity in case of SU(1,1) and classical interferometers for one-,two-, three- and four-photon absorption processes. The absorbance is found by conditioning the signal-to-noise ratio (\ref{eq.SNR}) becoming one, i.e. $\varepsilon_m / \Delta\varepsilon_m = 1$. 
}
\label{fig.new}
\end{figure*}

\section{Calculation of the scaling behaviour at large photon numbers}

Here we derive Eqs.~(\ref{eq.scalingDegrCoh}) and (\ref{eq.scaling}) of the main text. To this end, we first note that from Eq. \eqref{eq.LindbladNew} we have
\begin{align}
\mathcal{L}^{adj}_{mPA} [a^{'' \dagger}  a''] = - a^{'' \dagger  m} a^{'' m}. \label{eq.phot-number-change}
\end{align}

\subsection{Coherent state transmission measurements}

The variance of a coherent state with complex amplitude $\alpha$ is readily calculated as
\begin{align}
    \text{Var} (\hat{n}) &= \vert \alpha \vert^2
\end{align}
From Eq.~(\ref{eq.phot-number-change}), we get the change of the photon number
\begin{align}
    \frac{ \partial\langle \hat{n} \rangle}{\partial \varepsilon_m} &= - \vert \alpha\vert^{2m}. 
\end{align}
We next add internal and external losses, using the appropriate transformations above. In particular, the transformations~(\ref{eq.trafo-etaIn}) and (\ref{eq.trafo-etaEx}) effectively yield the replacement $\alpha \rightarrow \eta_i \alpha$. Consequently, we find $\text{Var} (\hat{n}) \rightarrow \eta_{In} \eta_{Ex} \text{Var} (\hat{n})$ and $\partial\langle \hat{n} \rangle / \partial \varepsilon_m \rightarrow  \eta_{In} \eta_{Ex} \partial\langle \hat{n} \rangle / \partial \varepsilon_m$. Altogether, we finally arrive at the result~(\ref{eq.scalingDegrCoh}) of the main text.

\subsection{Optimal states  in SU(1,1) measurements}

To find the photon number scaling of Eq.~(\ref{eq.sensitivity}), we first note that the second squeezing operation can be simplified in the limit of large $r_2$ to 
\begin{align}
    a''' &= \cosh (r_2) a'' + e^{i \phi_{Int}} \sinh (r_2)a''^\dagger \notag \\
    &\simeq \frac{1}{2} e^{r_2} \left( a'' +  e^{i \phi_{Int}} a''^\dagger \right) = \frac{ e^{r_2 + i \phi_{Int}/2}}{\sqrt{2}} \hat{q}_{\phi_{Int} /2}, \label{eq.2nd-squeezing}
\end{align}
where $\hat{q}_{\phi} = ( a'' e^{- i \phi} + a''^\dagger e^{i \phi} ) / \sqrt{2}$. 
With $\phi_{Int} = \pi$, this means the transformation turns any annihilation or creation operator into a momentum quadrature $\hat{p} = i (a^\dagger - a) / \sqrt{2}$ (modulo the prefactor) of the field at the sample. 
Therefore, the photon number variance of the light field at the detector is determined by the variance of the (squeezed) $\hat{p}$-quadrature at the sample, 
\begin{align}
    \text{Var} (\hat{n}''') &= \left(\frac{e^{2r_2}}{2}\right)^2 \text{Var} (\hat{p}''^2). \label{eq.Var_p}
\end{align}
To evaluate the latter, we write the operator as $\hat{p}'' = \Delta\hat{p} + \langle \hat{p} \rangle$. Here we have separated the expectation value (which grows with the photon number) from the squeezed fluctuation component $\Delta\hat{p} = \hat{p} - \langle \hat{p} \rangle$ with the properties $\langle \Delta\hat{p} \rangle = 0 $ and $\langle \Delta\hat{p}^2 \rangle = e^{- 2 r_1} / 2$. This allows us to write
\begin{align}
    \text{Var} (\hat{p}^2) &= \langle \hat{p}^4 \rangle - \langle \hat{p}^2 \rangle^2 \notag \\
    &= 4 \langle \hat{p} \rangle^2 \langle \Delta\hat{p}^2 \rangle + \left( \langle \Delta\hat{p}^4 \rangle - \langle \Delta\hat{p}^2 \rangle^2 \right).
\end{align}
The second term in this equation is again exponentially small, $\langle \Delta\hat{p}^4 \rangle - \langle \Delta\hat{p}^2 \rangle^2 = e^{-4r_1} / 2$. Consequently, the variance will be dominated by the first term, $4 \langle \hat{p} \rangle^2 \langle \Delta\hat{p}^2 \rangle$, unless the expectation value of $\hat{p}$ becomes very small. 

To continue, we note that we find \textit{numerically} for the cases $m = 1, \ldots 4$ that the optimal states approximately fulfill $\langle \hat{p} \rangle^2 \simeq \langle \hat{q}^2 \rangle = e^{2r_1} / 2$. Furthermore, we note that for the amplitude-squeezed state with $\phi_{Las} = \pi/2$ the expectation value of the squeezed quadrature is given by $\langle \hat{p} \rangle = \sqrt{2} \alpha e^{- r_1}$, which implies the optimal amplitude is given by $\alpha = e^{2 r_1} / 2$, and we find that, altogether,
\begin{align}
    \text{Var} (\hat{p}^2) &\simeq 4 \langle \hat{p} \rangle^2 \langle \Delta\hat{p}^2 \rangle = 1, \label{eq.OptVariance}
\end{align}
i.e. the variance of the optimal state remains constant for any photon number which is large enough to neglect the additional terms $\propto e^{- 4r_1}$. 

With this result, we simply need to show that the denominator in Eq. \eqref{eq.sensitivity} scales in the same way with the mean photon number as it did in the case of coherent states above, in order to prove the enhance sensitivity scaling. 
To this end, we first note that for the optimal state, the mean photon number reads 
\begin{align}
\langle \hat{n}\rangle + \frac{1}{2}&= ( \langle \hat{p}^2\rangle +\langle \hat{q}^2\rangle )/2 \simeq ( \langle \hat{p}\rangle^2 +\langle \hat{q}^2\rangle )/2 \notag \\
&\simeq \langle \hat{p}\rangle^2, \label{eq.phot-num}
\end{align}
where in the second line we used that $\langle \hat{q}^2\rangle = \langle \hat{p}\rangle^2$.
Next, we combine Eqs.~(\ref{eq.Lindblad-trafo}) and (\ref{eq.squeezing-trafo}) to obtain
\begin{align}
    \frac{\partial \langle\hat{n}\rangle}{\partial \varepsilon_m} &= \frac{e^{2r_2}}{2} \langle \mathcal{L}^{adj}_{mPA} [\hat{p}^2] \rangle \\
    &= \frac{e^{2r_2}}{2} \frac{1}{2} \left\langle\left( 2 \mathcal{L}^{adj}_{mPA} [a''^\dagger a''] - \mathcal{L}^{adj}_{mPA} [a''^2] - \mathcal{L}^{adj}_{mPA} [a''^{\dagger 2}] \right) \right\rangle \\
    &= \frac{e^{2r_2}}{4} \left\langle - 2 a''^{\dagger m} a''^m - \frac{m-1}{2} \left( a''^{\dagger m-2} a''^m + a''^{\dagger m} a''^{m-2} \right) \right\rangle. \label{eq.change_OptState}
\end{align}
If we assume for simplicity that these expectation values are dominated by the coherent amplitude (which for the optimal state is given by $\langle a \rangle = \sqrt{2} \langle \hat{p} \rangle$), we find that the first term in Eq.~(\ref{eq.change_OptState}) scales as $\sim \langle \hat{p} \rangle^{2m}$, whereas the other terms account only for subdominant scaling $\sim \langle \hat{p} \rangle^{2m-2}$. Hence, with Eq.~(\ref{eq.phot-num}), this analysis shows that the denominator in Eq.~(\ref{eq.sensitivity}) scales as $n_S^{m}$, just like in the measurement with coherent states analyzed above, but the variance~(\ref{eq.OptVariance}) remains constant. 
Taken together, this proves the optimal photon number scaling of the mPA sensitivity, as described by Eq.~(\ref{eq.scaling}) of the main text. 

\paragraph{External losses}
change the momentum operator to $\hat{p} \rightarrow \sqrt{\eta_{Ex}} \hat{p} + \sqrt{1-\eta_{Ex}} \hat{p}_2$ (where $p_2 = i (c_2^\dagger - c_2) / \sqrt{2}$ is the corresponding quadrature of the auxiliary mode). As a consequence, the variance~(\ref{eq.Var_p}) becomes 
\begin{align}
    \text{Var} (\hat{n}''') &\rightarrow \eta_{Ex}^2 \left(\frac{e^{2r_2}}{2}\right)^2 \text{Var} (\hat{p}''^2) + \eta_{Ex} (1-\eta_{Ex}) \frac{e^{2r_2}}{2} \left\langle\hat{p}^2 \right\rangle.
\end{align}
in the limit, where $r_2 \rightarrow \infty$, the second term in this equation becomes irrelevant. 
The denominator of Eq.~(\ref{eq.sensitivity}) simply becomes $\partial \langle \hat{n} \rangle / \partial\varepsilon_m = \eta_{Ex} e^{2r_2} \langle \mathcal{L}^{adj}_{mPA} [\hat{p}^2] \rangle / 2$, such that Eq.~(\ref{eq.sensitivity}) becomes independent of $\eta_{Ex}$. 

\paragraph{Internal losses} are not dealt with this easily. As we see numerically in Fig.~\ref{fig.results} for $m = 1, \ldots 4$, these change the optimal scaling. We did not find a simple analytical result to describe this behaviour. 

\subsection{Lindbladian approach to single-photon losses inside the sample}

In this section, we consider the scenario that single-photon losses take place inside the sample simultaneously with the weak mPA process, such that the time evolution of the propagating light field inside the sample is given by
\begin{align}
    \frac{d}{dt} \rho &= \left( \gamma_{mPA} \mathcal{L}_{mPA} + \gamma_{SPA} \mathcal{L}_{SPA}\right) \rho, \label{eq.mixed-evolution}
\end{align}
in which the first term accounts for mPA losses [see Eq.~(\ref{eq.Lindblad})], $\gamma_{SPA}$ is the single-photon loss coefficient and $\mathcal{L}_{SPA}$ describes single-photon loss inside the sample via the following Lindbladian 
\begin{align}
    \mathcal{L}_{SPA} \rho &= \left( 2 a \rho a^\dagger - a^\dagger a\rho - \rho a^\dagger a \right).
\end{align}
The solution to Eq. \eqref{eq.mixed-evolution} can be found straightforwardly as 
\begin{align}
    \rho (t) &= \exp \left( \mathcal{L}_{TPA} \varepsilon_{m} + \mathcal{L}_{SPA} \varepsilon_s \right) \rho_0,
\end{align}
in which $\varepsilon_s = \gamma_{SPA} \times t$. 
In our study, we investigated mPA perturbatively. The same can not be done for single-photon loss inside the sample as this process could be much larger. Considering that $\mathcal{L}_{mPA}$ and $\mathcal{L}_{SPA}$ do not commute, to evaluate the numerator and denominator of Eq. \eqref{eq.sensitivity}, we use the Suzuki-Trotter resulting in (see~\cite{Panahiyan2022})
\begin{align}
    &\quad \frac{\partial}{\partial\varepsilon_{m}} \exp \left( \mathcal{L}_{mPA} \varepsilon_{m} + \mathcal{L}_{SPA} \varepsilon_s \right) \\
    &= \int_0^1 \!\! dk \; \exp [ (1-k) \mathcal{L}_{SPA} \varepsilon_s ] \mathcal{L}_{mPA} \exp [ k \mathcal{L}_{SPA} \varepsilon_s],
\end{align}
which indicates that the single-photon loss inside the sample can be described by a convolution of losses taking place before and after the mPA event. We can use the beam splitter approach to model these losses with $\eta_{before} = \exp (-2 k \varepsilon_s)$ and $\eta_{after} = \exp (- 2 (1-k) \varepsilon_s)$ followed by a simple integral over $k$, and the total single photon loss is given by $\eta_{In} = \eta_{before} \eta_{after}= \exp (- 2 \varepsilon_s)$. This means that before the mPA transformation in Eq. \eqref{eq.Lindblad-trafo}, one should insert a beam splitter transformation $a' \rightarrow \sqrt{\eta_{before}} a' + \sqrt{1 - \eta_{before}} c_0$ to account for the single photon losses before the sample. The single-photon loss after the sample can be included into $\eta_{In}$ as it is pointed out before. Using the additional transformation for photon loss before the sample, one can find the variance and denominators of Eq. \eqref{eq.sensitivity} and take the integral with respect to $k$. Finding a closed expression analogous to Eq.~(\ref{eq.phot-number-change}) for the output field is difficult for general $m$ and arbitrary input states. However, we find the following behaviour for coherent state measurements with $2\leq m\leq 4$
\begin{align}
\Delta\varepsilon_{m, (coh)}^{2} &= \frac{(m-1)^{2}\log^2(\eta_{In})}{ (1-\eta_{In}^{m-1})^2}\frac{1}{\eta_{In}\eta_{Ex}n_{S}^{2m-1}}, \label{eq.NewCoh}
\end{align}
Comparing \eqref{eq.NewCoh} with \eqref{eq.scalingDegrCoh}, we notice that the first term in Eq. \eqref{eq.NewCoh} originates from the single-photon loss inside the sample which can highly degrade $\Delta\varepsilon_{m, (coh)}^{2}$. 
The full solution of the squeezed coherent state contains many different terms [see our previous discussion following Eq.~(\ref{eq.2nd-squeezing})], in general. If we only consider the change of the leading order in $n_s$ from Eq.~(\ref{eq.change_OptState}), we obtain the similar prefactor as for the coherent state. 

\subsection{The sensitivity in the limit of large second squeezing parameter and optimal first squeezing regime}

Here, we provide the explicit expressions for the variance $\Delta\varepsilon_{m, SU(1,1)}^2$ in the limit of large second squeezing parameter ($r_{2} \rightarrow \infty$) for one-, two-, three- and four-photon absorption processes. We can write it as
\begin{align}
\Delta\varepsilon_{m, SU(1,1)}^2 \bigg|_{r_{2} \rightarrow \infty} &= \frac{A}{\eta_{In}^2 B_{m}},
\end{align}
where 
\begin{align}
A=8 \eta _1 \left(2 \alpha ^2+2 \left(8 \alpha ^2+1\right) \eta _1 n_{\text{r1}}^2(1- \sqrt{n_{\text{r1}}^2+n_{\text{r1}}})-\left(4 \alpha ^2 \eta _1+4 \alpha ^2+1\right) \sqrt{n_{\text{r1}}^2+n_{\text{r1}}}+\left(\left(12 \alpha ^2+1\right) \eta _1+4 \alpha
   ^2+1\right) n_{\text{r1}}\right)+2, \label{AA}
\end{align}

\begin{align}
B_{1}=&  \left(4 \alpha ^2+1\right)^2 \left(2 n_{\text{r1}}-2 \sqrt{n_{\text{r1}}^2+n_{\text{r1}}}+1\right)^2,
\end{align}

\begin{align}
B_{2}=&  \bigg[\left(4 \alpha ^2+3\right) \alpha ^2+\left(32 \alpha ^4+48 \alpha ^2+6\right) \left[n_{\text{r1}}^2- \sqrt{n_{\text{r1}}^4+n_{\text{r1}}^3} \right]+4 \left(8 \alpha ^4+10 \alpha ^2+1\right) n_{\text{r1}}-\left(16 \left(\alpha ^4+\alpha ^2\right)+1\right) \sqrt{n_{\text{r1}}^2+n_{\text{r1}}}\bigg]^2,
\end{align}

\begin{align}
B_{3}=&  \bigg[\left(4 \alpha ^2+5\right) \alpha ^4-2 \left(12 \alpha ^4+23 \alpha ^2+6\right) \alpha ^2 \sqrt{n_{\text{r1}}^2+n_{\text{r1}}}+3 \left(4 \alpha^2+3\right) \left(16 \alpha ^4+40 \alpha ^2+3\right) n_{\text{r1}}^2+\left(72 \alpha ^6+178 \alpha^4+78 \alpha ^2+3\right) n_{\text{r1}}  \notag
\\
&
+2 \left(64 \alpha ^6+240 \alpha ^4+180 \alpha ^2+15\right) \left[ n_{\text{r1}}^3-\sqrt{n_{\text{r1}}^6+n_{\text{r1}}^5}\right]-4 \left(32 \alpha ^6+96 \alpha   ^4+54 \alpha ^2+3\right) \sqrt{n_{\text{r1}}^4+n_{\text{r1}}^3}\bigg]^2,
\end{align}

\begin{align}
B_{4}= & \bigg[\left(4 \alpha ^2+7\right) \alpha ^6-\left(32 \alpha ^4+92 \alpha ^2+45\right) \alpha ^4 \sqrt{n_{\text{r1}}^2+n_{\text{r1}}}+\left(128 \alpha ^6+488\alpha ^4+414 \alpha ^2+63\right) \alpha ^2 n_{\text{r1}}+  \notag
\\&
2 \left(256\alpha ^8+1792 \alpha ^6+3360 \alpha ^4+1680 \alpha ^2+105\right)\bigg[ n_{\text{r1}}^4-\sqrt{n_{\text{r1}}^8+n_{\text{r1}}^7}\bigg]+  \notag
\\&
\left(640 \alpha ^8+3352 \alpha ^6+4458 \alpha ^4+1467 \alpha ^2+54\right) n_{\text{r1}}^2
+16 \left(64 \alpha ^8+400 \alpha ^6+660 \alpha ^4+285 \alpha ^2+15\right) n_{\text{r1}}^3-  \notag
\\&
3 \left(256 \alpha ^8+1536 \alpha ^6+2400 \alpha ^4+960 \alpha^2+45\right) \sqrt{n_{\text{r1}}^6+n_{\text{r1}}^5}-\left(320 \alpha^8+1496 \alpha ^6+1698 \alpha ^4+432 \alpha ^2+9\right) \sqrt{n_{\text{r1}}^4+n_{\text{r1}}^3}\bigg]^2,  \label{BB}
\end{align}
in which for the sake brevity, we have used $\sinh^2(r_{1})=n_{\text{1}}$

Evidently, these expressions are independent of external loss. While we are using the above formulas to conduct our study, it should be noted that threshold for $\Delta\varepsilon_{2, SU(1,1)}^2$ converging to $\Delta\varepsilon_{2, SU(1,1)}^2 \bigg|_{r_{2} \rightarrow \infty}$ 
can be reached with realistic squeezing strengths.
In the panel (a) of Fig. \ref{fig.new}, we show for $m=2$ numerically how the variance~(\ref{eq.sensitivity}) converges to the $r_{2} \rightarrow \infty$-limit with increasing second squeezing parameter, and for several values of the photon number at the sample $n_S$. 
The convergence between $\Delta\varepsilon_{2, SU(1,1)}^2$ and $\Delta\varepsilon_{2, SU(1,1)}^2 \bigg|_{r_{2} \rightarrow \infty}$ is dictated by three factors - the mean photon number, internal loss, and the degree of the nonlinearity.
We should highlight that similar numerical evaluation can be done for arbitrary $m$ processes and the results would be the same. This regime of convergence for second squeezing parameter can be experimentally realized and it is to this converging behaviour which allows us to use Eqs. \ref{AA}-\ref{BB} to investigate the enhancement in SU(1,1) interferometers in the high-gain regime of $r_{2} \rightarrow \infty$.  

Using the limit $r_{2} \rightarrow \infty$ (Eqs. \ref{AA}-\ref{BB}), the next step is finding the optimum sensitivity by optimizing the first squeezing parameter for different internal loss and incident photon numbers. In the panel (b) of Fig. \ref{fig.new}, we have done so for a TPA process in the case of different internal losses ($\eta_{In}=1$ and $0.9$). The white area in these diagrams corresponds to unphysical parameters, where $\sinh^2 (r_1) > n_S$.
The border line between white area and contour plots is the (unseeded) squeezed vacuum case, i.e. $r_{1} = \text{arcsinh} \sqrt{n_{S}}$ and $\alpha = 0$. Within the allowed region for the first squeezing parameter, there is a unique optimal first squeezing parameter at each incident photon number and internal loss. This unique first squeezing parameter is an increasing function of the incident photon number and a decreasing function of the internal loss. In the absence of internal losses, we find that it satisfies approximately $\langle \hat{p} \rangle^2 = \langle \hat{q}^2 \rangle$, as we used in our derivation above.  

We can use the obtained optimal regime of the first squeezing parameter to 
investigate the minimal number of photons $n_S^{min}$ necessary to observe a signal. 
This minimal photon number is where the signal to noise ratio (SNR)
\begin{align}
SNR= \varepsilon_m\frac{\left| \frac{\partial \langle \hat{n} \rangle}{ \partial \varepsilon_m } \right| }{\sqrt{\text{ Var } (\hat{n})}}=\frac{ \varepsilon_m}{\Delta\varepsilon_{m }}, \label{eq.SNR}
\end{align}
becomes one, i.e. the signal becomes as large as the noise. 
In panel (c) of Fig. \ref{fig.new}, we have plotted absorbance as a function of the minimal photon number for  SU(1,1) interferometer and classical measurements with different degrees of mPA and internal loss. In an ideal case, the unit SNR would be achieved at a considerably smaller photon number in an SU(1,1) interferometer compared to its classical counterpart for the same absorbance. Although the enhancement offered by interferometric detection can be negatively affected by internal loss, even in the presence of strong internal loss, our strategy still outperforms any classical measurement. It should be noted that external loss would degrade the results for the conditioned SNR of the classical strategy even further whereas our strategy in SU(1,1) interferometer is independent of external loss.

\section{The relation to absorption cross sections and estimate of their magnitude}

Here we discuss the relation between the absorbance $\varepsilon_m$ and experimentally measured absorption cross sections. 

To this end, we first turn to the measurement of two-photon cross sections, and make the connection to the measurement of TPA with squeezed vacuum, in particular.
The measurement is typically related to the change of the photon count rate $\Delta R_2$ with and without the sample inserted in the beam. Taking a squeezed vacuum input state, this change is then related to the photon flux density $\phi$ as
\begin{align}
    \Delta R_2 &= \sigma_e \phi + \delta_r \phi^2.
\end{align}
Here, we have used the conventional nomenclature, where $\sigma_e$ (in units $m^{2}$) denotes the entangled absorption cross section stemming from photon pairs that are generated in the same downconversion event. $\delta_r$ ($m^4 s$) is the classical two-photon absorption cross section,  the corresponding change of the photon count rate scales quadratically with the photon flux density. 

We next follow the arguments by Parzuchowski et al.~\cite{Parzuchowski2021}, and relate the photon flux density to the mean photon number of a single mode state (per pulse) as
\begin{align}
    \phi = \frac{ \langle \hat{n} \rangle }{ T A }, 
\end{align}
where we define the pulse duration $T$ and the beam area $A$. 
We calculate the change of the photon number using the formalism presented in the previous sections. 
For a squeezed vacuum, this yields in the absence of single-photon losses $\partial \langle\hat{n}\rangle / \partial\varepsilon_2 = n_0 + 3n_{0}^{2}$ ~\cite{Panahiyan2022}, where $n_0 = \sinh^2 (r)$ is the mean photon number of the pulse (For a coherent state with complex amplitude $\alpha$, we would likewise obtain $\partial \langle\hat{n}\rangle / \partial\varepsilon_2 = n_{\alpha}^{2}$ with mean photon number $n_{\alpha} = |\alpha|^2$). 
We further assume that two-photon absorption takes place in a sample with density $\nu$ and volume $V = A \ell$, where $\ell$ is the sample length. 
We can then identify the change of the photon count rate as $\Delta R_2 = \varepsilon_2 \partial \langle\hat{n}\rangle/ \partial\varepsilon_2 T^{-1}$, assuming that the photon flux across the full beam area $A$ is detected. 
Taken together, we can use these relations to turn the absorbance into a cross section by writing
\begin{align}
    \varepsilon_2 = \sigma_e \nu \ell.
\end{align}
Likewise, we can connect the entangled cross section to the classical counterpart as
\begin{align}
    \delta_r = 3 \sigma_e T A,
\end{align}
$i.e.$ the classical two-absorption cross section will in general depend on the experimental parameters. 

We next provide an estimate for the absorbance one could expect in an experiment. We use parameters from Ref.~\cite{Villabona17} for a two-photon absorption measurement in rhodamine B molecules, $\sigma_e = 4.2 \times 10^{-18} \; \text{cm}^2$ / molecule and a concentration of $38 \; \mu M$. Assuming further a reasonable length of a cuvette of $\ell = 1 \; mm$, we end up with $\varepsilon_2 \simeq 10^{-2}$. We remark that the values of entangled two-photon absorption cross sections vary strongly in the literature (for a review of the debate, see e.g. \cite{Hickam2022}), and we employed one of the largest reported cross sections in our estimate. Hence, $\varepsilon_2 \simeq 10^{-2}$ should perhaps be considered an upper bound for realistic absorbances. 

To the best of our knowledge, multi-photon absorption of squeezed light has not been reported beyond $m = 2$ to date. However, we can obtain a rough estimate for $\varepsilon_m$ by noting that in semiclassical light-matter interactions the $m$-photon absorption rate is related to the $2m -1$-th nonlinear susceptibility~\cite{CRONSTRAND2005}. Its strength can be estimated according to the argument provided in Boyd's book~\cite{Boyd} relating the nonlinear susceptibility to the characteristic atomic field strength $E_{at} = 5.14 \times 10^{11}$~V / m. Thus, we have $\chi^{(n)} \sim \chi^{(1)} / E_{at}^{n-1}$, where the linear susceptibility $\chi^{(1)} = \mathcal{O} (1)$ is the linear susceptibility. 
Counteracting this steep decrease, we can expect an enhancement of nonlinear light-matter interactions due to the photon statistics of the light fields related to its normally ordered, normalized correlation function $g^{(m)}$ and the $m$-th power of the photon number, $\langle \hat{n} \rangle^m$ \cite{Mollow, Agarwal1970}, which for the case of squeezed vacuum increases as a double-factorial $\sim (2 m -1)!!$ \cite{Spasibko2017}. We therefore end up with
\begin{align}
    \varepsilon_m \sim (2m-1)!! \times \langle \hat{n} \rangle^m / E_{at}^{2m-2}.
\end{align}
This scaling behaviour does not enable us to obtain explicit estimates for $\varepsilon_m$, but as long as the involved field intensities (corresponding to $\langle\hat{n}\rangle$) are much smaller than $E_{at}^2$, i.e. as long as the light-matter interaction can be treated perturbatively, they can be expected to decrease rapidly with $m$. 

\end{widetext}

\end{document}